\documentclass[twocolumn,secnumarabic,amssymb, amsmath, nofootinbib,tightenlines, nobibnotes, aps, prl,epsfig]{revtex4}
\usepackage{graphicx}% Include figure files
\usepackage{dcolumn}% Align table columns on decimal point
\usepackage{bm}% bold math
\begin{document}
\preprint{APS/123-QED}
\title{ Nonlinear corrections on the parametrization methods}% Force line breaks with \\

\author{G.R.Boroun}%
 \email{grboroun@gmail.com; boroun@razi.ac.ir }
\author{B.Rezaei }
\altaffiliation{brezaei@razi.ac.ir}%Lines break automatically or can be forced with \\

\affiliation{Department of  Physics, Razi University, Kermanshah
67149, Iran}% \textbackslash\textbackslash

%\author{D.Schildknecht}%
%\email{schild@physik.uni-bielefeld.de } \affiliation{
%Fakult$\ddot{a}$t f$\ddot{u}$r Physik, Universit$\ddot{a}$t
%Bielefeld,
%D-33501 Bielefeld, Germany}% \textbackslash\textbackslash
\date{\today}% It is always \today, today,
\begin{abstract}
%%%%%%%%%%%%%%%%%%%%%%%%%%%%%%%%%%%%%%%%%%%%%%%%%%%%%%%
We present non-linear corrections (NLC) to the distribution
functions at low values of $x$ and $Q^{2}$ using the
parametrization $F_{2}(x,Q^{2})$ and $F_{L}(x,Q^{2})$. We use a
direct method to extract non-linear corrections to the ratio of
structure functions and the reduced cross section  in the
next-to-next-to-leading order (NNLO) approximation with respect to
the  parametrization method (PM). Comparison between the
non-linear results with the bounds in color dipole model (CDM) and
HERA data indicate the consistency of the non-linear behavior of
the gluon distribution function at low-$x$ and low-$Q^{2}$. The
non-linear longitudinal structure functions are comparable with
the H1 Collaboration data in a wide range of $Q^{2}$ values.
Consequently, the non-linear corrections at NNLO approximation to
the reduced cross sections at low and moderate $Q^{2}$ values
 show  good agreement with the HERA combined
data. These results at low $x$ and low $Q^{2}$ can be
applied to the LHeC region for analyses of ultra-high-energy processes.\\
%%%%%%%%%%%%%%%%%%%%%%%%%%%%%%%%%%%%%%%%%%%%%%%%%%%%%%%
\end{abstract}
 \pacs{***}%PACS, the Physics and Astronomy
                              %Classification Scheme.
\keywords{****} %Use showkeys class option if keyword
\maketitle
%**********************************************************
%%%%%%%%%%%%%%%%%%%%%%%%%%%%%%%%%%%%%%%%%%%%%%%%%%%%%%%%%%%%%%%%%%%%%%%%%%%%%%%%%%%%%%%%%%%%%
\subsection{1. Introduction}

Based on parton model and perturbative quantum chromodynamics
(pQCD), the Dokshitzer-Gribov-Lipatov-Altarelli-Parisi (DGLAP)
evolution equations [1] successfully and quantitatively interpret
the $Q^{2}$-dependence of parton distribution functions (PDFs).
The small-$x$ behavior due to DGLAP equations is driven by input
distributions at a starting scale $Q=Q_{0}$. In recent years
several different parametrizations of PDF from a global fits to
the available data have been introduced [2-4]. PDF groups, such as
Refs.[2,3], analysis HERA data [5] in their global analysis.
Recently in Ref.[4] authors presented the parton distribution
functions using a wide variety of high-precision Large Hadron
Collider (LHC) data in addition to the combined HERA I+II
deep-inelastic scattering data set. LHC covered data sample of
over $140~ \mathrm{fb}^{-1}$ at the $13~ \mathrm{TeV}$ run for
both ATLAS and CMS collaborations. The CT global analyses [4]
explored a broad range of parametric forms for the parton
distribution functions at the starting scale, $Q=Q_{0}$. In CT18
the initial nonperturbative parametrizations are written in the
following formal form
$$
f_{i}(x,Q_{0})=\alpha_{0}x^{\alpha_{1}-1}(1-x)^{\alpha_{2}}P_{i}(y,\alpha_{3},\alpha_{4},...),
$$
where the coefficients $\alpha_{1}$ and $\alpha_{2}$ control the
asymptotic behavior of $f_{i}(x,Q_{0})$ in the limits
$x{\rightarrow}0$ and $1$, and $P_{i}$ is a sum of Bernstein
polynomials dependent on $y=f(x)$ which is very
flexible across the whole interval $0<x<1$.\\
At extremely small of the Bjorken variable $x$,  the pQCD
evolution provides a rather singular behavior of the PDFs which
strongly violates the Froissart boundary [6]. With respect to HERA
data, some new parametrizations of the proton structure function
have been proposed by authors in Refs.[7,8] in a wide range of
$Q^{2}$ values which are in a full accordance with the Froissart
predictions. These parametrizations are relevant in investigations
of ultra-high energy processes, such as scattering of cosmic
neutrino off hadrons. The importance of ultra-high energies is in
explore extreme regions of the $(Q^{2}, x)$ phase space, where
non-accelerator data exist [9]. Studies along energy boost not
only confirm HERA investigations but also provide crucial
benchmarks for further investigations of the high-energy limit of
QCD at the Large Hadron electron Collider (LHeC) [10]. The
kinematic extension of the LHeC will allow us to examine the
non-linear dynamics at low $x$ [11,12]. The non-linear region is
approached when the reaction is mediated by multi-gluon exchange.
Indeed the gluon -gluon recombination processes cause that the
growth of the gluon density is slowed down at smaller values of
$x$ and $Q^{2}$ (but still $Q^{2}{\gg}\Lambda^{2}_{QCD}$) [13,14].
At this region gluon recombination terms, which lead to non-linear
corrections to the evolution equations, can become significant.
Gluon recombination in deep inelastic and diffractive scattering
were published some time ago in Refs.[14] and [15] respectively.
The gluon density cannot grow forever because hadronic
cross-sections comply with the unitary bound known as Froissart
Bound [6]. Indeed the unitarity (or Froissart bound states) does
not grow faster than $\sigma<\pi d^{2}\ln^{2}(s/s_{0})$ where $d$
is some typical hadronic scale and  $s$ is the Mandelstam variable
denoting the square of the total invariant energy of the process.
The gluon recombination effects tamed growth of the gluon density
towards low $x$. These effects induce non-linear power corrections
to the DGLAP equations. Some studies of the non-linear behavior of
PDFs are given in Refs.[16-30] in recent
years.\\
Low $x$ physics at the LHeC and Future Circular Collider
hadron-electron (FCC-he) is an area for discovery non-linear
effects. This extends the kinematic reach, of maximum
$Q^{2}{\simeq}1~\mathrm{TeV}^{2}$ and
$x{\propto}Q^{2}/s{\simeq}~10^{-5...-6}$ for LHeC and $10^{-7}$
for FCC-he , where parton interaction must to become non-linear.
The extended kinematic range of the LHeC provides unique avenues
to explore the possible onset of non-linear QCD dynamics at
small-$x$ [10,11,12]. In Ref.[31] PDF4LHC15  includes HERA data
down to $x{\simeq}~10^{-4}$ which is successfully described via
the DGLAP framework.\\
The non-linear evolution in high-density QCD is considered in a
shock wave color field of the target in Ref.[32]. Authors have
considered deeply inelastic scattering at very high energies in
the saturation regime and have developed a formalism which allows
to evaluate successively the non-linearities in the generalized
evolution equation for the dipole densities. In Ref.[33] authors
have discussed the results of the analytical and numerical
analysis of the non-linear Balitsky-Kovchegov equation. One of the
important outcomes of studied in Ref.[33] is the existence of the
saturation scale $Q_{s}(x)$ which is a characteristic scale at
which the parton recombination effects become important. The
solution to the non-linear equation has the property of the
geometric scaling in the regime where $k<Q_{s}(x )$ whereas in the
case when $k>Q_{s}(x)$ the solution enters the linear regime,
where $k$ is the gluon transverse momenta. The phenomenological
implications of the parton distribution functions sets with
small-$x$ resummation for the longitudinal structure function
$F_{L}$ at HERA have investigated in Ref.[34]. Also a resolution
of the incorporating $\ln(1/x)$-resummation terms into the HERAPDF
fits have investigated using the xFitter program. Authors in
Ref.[35] have tried to investigate solutions of the non-linear
evolution equation in the nonperturbative part of the low $x$
region.\\
The non-linear terms have been calculated by Gribov-Levin- Ryskin
(GLR) and Mueller-Qiu (MQ) in [36]. GLR originally showed how to
qualitatively modify the DGLAP gluon evolution equation in order
to incorporate effects of gluon recombination, then MQ derived the
singlet evolution equation for the conversion of gluon to quarks.
The modified evolution equations, due to the fusion of two gluon
ladders, denote as the GLR-MQ equations
\begin{eqnarray}
\frac{\partial{xg(x,Q^{2})}}{\partial{\ln}Q^{2}}&=&\frac{\partial{xg(x,Q^{2})}}{\partial{\ln}Q^{2}}|_{DGLAP}\nonumber\\
&&-\frac{81}{16}\frac{\alpha_{s}^{2}(Q^{2})}{\mathcal{R}^{2}Q^{2}}\int_{\chi}^{1}\frac{dz}{z}[\frac{x}{z}g(\frac{x}{z},Q^{2})]^{2},
\end{eqnarray}
 and
\begin{eqnarray}
\frac{\partial{F_{2}(x,Q^{2})}}{\partial{\ln}Q^{2}}&=&\frac{\partial{F_{2}(x,Q^{2})}}{\partial{\ln}Q^{2}}|_{DGLAP}\nonumber\\
&&-\frac{5}{18}\frac{27\alpha_{s}^{2}(Q^{2})}{160\mathcal{R}^{2}Q^{2}}[xg(x,Q^{2})]^{2},
\end{eqnarray}
where $g(x,Q^{2})$ is the gluon density, $xg(x,Q^{2})=G(x,Q^{2})$
is the gluon density momentum. Here $\chi=\frac{x}{x_{0}}$ and
$x_{0}$ is the boundary condition that the gluon distribution
(i.e., $G(x,Q^{2})$) joints smoothly onto the linear region. The
correlation length $\mathcal{R}$  determines the size of the
non-linear terms. This value depends on how the gluon ladders are
coupled to the nucleon or on how the gluons are distributed within
the nucleon. The $\mathcal{R}$ is approximately equal to $\simeq
5~\mathrm{GeV}^{-1}$ if the gluons are populated across the proton
and it is equal to $\simeq 2~\mathrm{GeV}^{-1}$ if the gluons have
hotspot like structure. Here the higher dimensional gluon
distribution(i.e., higher twist) is assumed to be zero. In the
small $x$ region, the saturation scale $Q_{s}^{2}(x)$
($Q_{s}^{2}=Q_{0}^{2}(x/x_{0})^{-\lambda}$ where $Q_{0}$ and
$x_{0}$ are free parameters) indicates the saturation limit where
the DGLAP and GLR-MQ terms in the non-linear equation become equal
and is usually defined as [9,37]
\begin{eqnarray}
\frac{\partial{xg(x,Q^{2})}}{\partial{\ln}Q^{2}}|_{Q^{2}=Q^{2}_{s}(x)}=0~\mathrm{and}
~\frac{\partial{xS(x,Q^{2})}}{\partial{\ln}Q^{2}}|_{Q^{2}=Q^{2}_{s}(x)}=0,\nonumber
\end{eqnarray}
where $xS(x,Q^{2})$ is the sea quark distribution. At
$Q^{2}>Q^{2}_{s}(x)$ the linear DGLAP part in the region of
applicability of the DGLAP+GLRMQ is dominant and at
$Q^{2}<Q^{2}_{s}(x)$ the GLRMQ terms dominant as all non-linear
terms become important, or equivalently
\begin{eqnarray}
\frac{\mathrm{non-linear}~ \mathrm{terms}}{\mathrm{linear}~
\mathrm{terms}}|_{Q^{2}=Q^{2}_{s}(x)}=1.\nonumber
\end{eqnarray}
This balances the linear and non-linear splitting effects at
$Q^{2}=Q^{2}_{s}(x)$.\\
 This paper is organized as follows. In the next section the
 theoretical formalism is presented, including the non-linear
 evolution and the  parametrization models.  In section 3, we present
  a detailed numerical analysis and our main results. We then confront these
results with  the CDM bounds and the HERA data at low values of
$Q^{2}$. In the last
section  we summarize our main conclusions and remarks.\\
%%%%%%%%%%%%%%%%%%%%%%%%%%%%%%%%%%%%%%%%%%%%%%%%%%%%%%%%%%%%%%%%%%

\subsection{2. Theoretical formalism}

The structure function $F_{2}$ is expressed through the quark and
gluon densities as
\begin{eqnarray}
F_{2}(x,Q^{2})&=&B_{2,s}(x,Q^{2}){\otimes}F_{2}^{s}(x,Q^{2})\nonumber\\
&&+<e^{2}>B_{2,g}(x,Q^{2}){\otimes}xg(x,Q^{2}),
\end{eqnarray}
where  $B_{2,a}(a=s,g)$ are the common Wilson coefficient
functions and the symbol ${\otimes}$ denotes a convolution
according to the usual prescription,
$f(x){\otimes}g(x)=\int_{x}^{1}\frac{dy}{y}f(y)g(\frac{x}{y})$.
Here  using the fact that the non-singlet contribution can be
ignored safely at low values of $x$. The  DGLAP evolution
equations can be written as
%\begin{widetext}
\begin{eqnarray}
\frac{{\partial}F_{2}^{s}(x,Q^{2})}{{\partial}{\ln}Q^{2}}&=&-\frac{a_{s}(Q^{2})}{2}[(P_{ss}^{(0)}(x)
+a_{s}(Q^{2})\widetilde{P}_{ss}^{(1)}(x)\nonumber\\
&&+...){\otimes}F_{2}^{s}(x,Q^{2}) +<e^{2}>(P_{sg}^{(0)}(x)\nonumber\\
&&+a_{s}(Q^{2})\widetilde{P}_{sg}^{(1)}(x)
+...){\otimes}xg(x,Q^{2})],\nonumber\\
\frac{{\partial}xg(x,Q^{2})}{{\partial}{\ln}Q^{2}}&=&-\frac{a_{s}(Q^{2})}{2}[<e^{2}>^{-1}(P_{gs}^{(0)}(x)
+a_{s}(Q^{2})\nonumber\\
&&{\times} \widetilde{P}_{gs}^{(1)}(x)
+...){\otimes}F_{2}^{s}(x,Q^{2})\nonumber\\
&&+(P_{gg}^{(0)}(x) +a_{s}(Q^{2})\widetilde{P}_{gg}^{(1)}(x)\nonumber\\
&&+...){\otimes}xg(x,Q^{2})],
\end{eqnarray}
%\end{widetext}
where
\begin{eqnarray}
\widetilde{P}_{ab}^{(n)}(x)={P}_{ab}^{(n)}(x)+[B_{2,s}+B_{2,g}+...]\otimes
{P}_{ab}^{(0)}(x)+... .\nonumber
\end{eqnarray}
The quantities $\widetilde{P}_{ab}$$^{,}s$ are expressed via the
known splitting  and Wilson coefficient functions  in literatures
[38,39] and $P_{ab}$ are the splitting functions in [40]. The
running coupling  in the high-loop corrections of the above
equation is expressed entirely  thorough the variable
$a_{s}(Q^{2})$ where
$a_{s}(Q^{2})=\frac{\alpha_{s}(Q^{2})}{4\pi}$.  Also $<e^{k}>$ is
the average of the charge $e^{k}$ for the active quark flavors,
$<e^{k}>=n_{f}^{-1}\sum_{i=1}^{n_{f}}e_{i}^{k}$.\\
In perturbative QCD, the longitudinal structure function in terms
of the coefficient functions at small $x$ is given by [41]
\begin{eqnarray}
F_{L}(x,Q^{2})&=&C_{L,q}(\alpha_{s},x){\otimes}F_{2}^{s}(x,Q^{2})\nonumber\\
&&+<e^{2}>C_{L,g}(\alpha_{s},x){\otimes}xg(x,Q^{2}),
\end{eqnarray}
where  the coefficient functions can be written as [42]
\begin{eqnarray}
C_{L,a}(\alpha_{s},x)=\sum_{n=1}a_{s}(Q^{2})^{n}c_{L,a}^{n}(x),
\end{eqnarray}
and $n$ is the order in the running coupling.\\
The proton structure function parametrized in Refs.[7] and [8]
with a global fit function  to the ZEUS data  and to combined HERA
data respectively for $F_{ 2}(x,Q^{2})$ in a wide range of $Q^{2}$
at $x< 0.1$, as
\begin{eqnarray}
F_{ 2} (x,Q^{2})& =& (1- x)[\frac{F_{P}}{1 -x_{P}} +
A(Q^{2})\ln( \frac{x_{P}}{ x}\frac{1 - x}{ 1 - x_{P}})\nonumber\\
&&+B(Q^{2}) \ln^{2}( \frac{x_{P}}{ x}\frac{1 - x}{ 1 - x_{P}})].
\end{eqnarray}
Here $x_{P}$ is an approximate fixed point observed in the data
where curves of $F^{p}_{2}(x,Q^{2})$ for different $Q^{2}$ cross.
Also
\begin{eqnarray}
 A(Q^{2}) = a_{0} + a_{1} {\ln}Q^{2} + a_{2} {\ln}^{2}
 Q^{2},\nonumber
 \end{eqnarray}
and
\begin{eqnarray}
  B(Q^{2}) = b_{0} + b_{1}
{\ln}Q^{2} + b_{2} {\ln}^{2} Q^{2}.\nonumber
\end{eqnarray}
The fitted parameters are tabulated in Table I. In terms of the
measured structure function $F_{2}(x,Q^{2})$ (i.e., Eq.(7)), the
gluon distribution function is determined due to the Laplace
transform method in Ref.[7]. In the case of four massless quarks,
the gluon distribution function is obtained with an expression
quadratic in both $\ln{Q^{2}}$ and $\ln{(1/x)}$ for $0<x ~{\leq}~
0.06$ as
\begin{eqnarray}
G(x,Q^{2})& =&
\frac{3}{5}[-2.94-0.359~{\ln}Q^{2}-0.101~{\ln}^{2}Q^{2}\nonumber\\
&&+(0.594-0.0792~{\ln}Q^{2}-0.000578~{\ln}^{2}Q^{2})\nonumber\\
&&{\times}\ln(1/x)+(0.168+0.138~{\ln}Q^{2}\nonumber\\
&&+0.0169~{\ln}^{2}Q^{2})\ln^{2}(1/x)].
\end{eqnarray}
In Refs.[43,44] the behavior of the longitudinal structure
function due to the Mellin transform and Regge theory have been
considered respectively. The authors in Ref.[43] obtained
analytical relations for the longitudinal structure function at LO
and NLO approximations in terms of the effective parameters of the
parametrization of the proton structure function. With respect to
the Mellin transform method, the LO and NLO longitudinal structure
functions are obtained at low $x$ by the following forms
\begin{eqnarray}
F_{L}^{\mathrm{LO}}(x,Q^{2})& =&(1-
x)^{n}\sum_{m=0}^{2}C_{m}(Q^{2})L^{m},
\end{eqnarray}
where $L^{,}$s are the logarithmic terms, and
\begin{eqnarray}
F_{L}^{\mathrm{NLO}}(x,Q^{2})&=&\frac{1}{[1+\frac{1}{3}a_{s}(Q^{2})L_{C}
(\widehat{\delta}^{(1)}_{sg}-\widehat{R}^{(1)}_{L,g})]}\bigg{\{}\nonumber\\
&&[1-a_{s}(Q^{2})
(\overline{\delta}^{(1)}_{sg}-\overline{R}^{(1)}_{L,g})]F_{L}^{\mathrm{LO}}(x,Q^{2})\nonumber\\
&&-a^{2}_{s}(Q^{2})[\frac{1}{3}\widehat{B}^{(1)}_{L,s}L_{A}+\overline{B}^{(1)}_{L,s}]F_{2}(x,Q^{2})
\bigg{\}}.\nonumber\\
\end{eqnarray}
The coefficient functions at LO and NLO approximations are
summarized in Appendix A and also the effective parameters are
defined in Table II. Therefore the parametrization of
$\sigma_{r}(x,Q^{2})$ in terms of the Froissart-bounded
parametrizations of $F_{2}(x,Q^{2})$ and $F_{L}(x,Q^{2})$  at LO
and NLO approximations read as
\begin{eqnarray}
\sigma_{r}^{(n)}(x,Q^{2})&=&D(Q^{2})(1-x)^{\nu}\sum_{m=0}^{2}A_{m}(Q^{2})L^{m}\nonumber\\
&&-f(y)F_{L}^{(n)}(x,Q^{2})
\end{eqnarray}
where $f(y)={y^2}/{Y_{+}}$, $Y_{+}=1+(1-y)^2$ and $y=Q^{2}/{sx}$.
The first and second order results (i.e., $n=1$ and $2$) shows the
LO ($n=1$) and NLO ($n=2$) longitudinal coefficient functions.
Recently, the non-linear modification of the evolution of the
gluon density at small $x$ is considered in the leading order of
perturbation theory in Ref.[45]. In Ref.[17] the important role of
absorptive effects and power corrections in low $x$ DGLAP
evolution are considered. These effects flat the behavior of the
low $x$ gluon density which arises from the freezing of
$\alpha_{s}$ at low $Q^{2}$ values. In the following the
extraction of the non-linear corrections provides means for
determining distribution functions and reduced cross sections at
low $x$ and low $Q^{2}$ values with respect to the
phenomenological assumptions. In the following, these non-linear
modifications will be applied
to the distribution functions at high-order corrections.\\
%%%%%%%%%%%%%%%%%%%%%%%%%%%%%%%%%%%%%%%%%%%%%%%%
\subsection{3. Results and discussions}

Effects of non-linear gluon corrections are obtained by solving
the GLR-MQ equation (i.e., Eq.(1)) in standard form
\begin{eqnarray}
\frac{\partial{G(x,Q^{2})}}{\partial{\ln}Q^{2}}&=&\frac{\partial{G(x,Q^{2})}}{\partial{\ln}Q^{2}}|_{DGLAP}\nonumber\\
&&-\frac{81}{16}\frac{\alpha_{s}^{2}(Q^{2})}{\mathcal{R}^{2}Q^{2}}\int_{\chi}^{1}\frac{dz}{z}G^{2}(\frac{x}{z},Q^{2}).
\end{eqnarray}
By solving the above equation (i.e., Eq.(12)), the nonlinear
corrections to the gluon distribution function (i.e.,
$G^{\mathrm{NLC}}(x,Q^{2})$ ) is obtained by the following form as
\begin{eqnarray}
G^{\mathrm{NLC}}(x,Q^{2})&=&G^{\mathrm{NLC}}(x,Q_{0}^{2})+[G(x,Q^{2})-G(x,Q_{0}^{2})]\nonumber\\
&&-\int_{Q_{0}^{2}}^{Q^{2}}\frac{81}{16}\frac{\alpha_{s}^{2}(Q^{2})}{\mathcal{R}^{2}Q^{2}}\int_{\chi}^{1}\frac{dz}{z}G^{2}(\frac{x}{z},Q^{2})d{\ln}Q^{2}\nonumber\\
\end{eqnarray}
where $G(x,Q^{2})$ and $G(x,Q_{0}^{2})$ are the unshadowed gluon
distributions and obtained from the solutions to standard DGLAP
equations which determined through a fit to HERA data (according
to Eq.(8)). We note that at $x{\geq}x_{0}(=10^{-2})$ the
non-linear corrections are negligible. At the initial scale
$Q_{0}^{2}$, the low $x$ behavior of the non-linear gluon
distribution is assumed to be [37]
\begin{eqnarray}
G^{\mathrm{NLC}}(x,Q_{0}^{2})&=&G(x,Q_{0}^{2})\{1+\frac{27\pi{\alpha_{s}(Q_{0}^{2})}}{16\mathcal{R}^{2}Q_{0}^{2}}\theta(x_{0}-x)\nonumber\\
&&{\times}[G(x,Q_{0}^{2})-G(x_{0},Q_{0}^{2})] \}^{-1}.
\end{eqnarray}
Indeed authors in Ref.[37] impose shadowing corrections by
modifying gluon density for $x<x_{0}$, where the leading shadowing
approximation $g_{sat}$ is the value of the gluon which would
saturate the unitarity limit as
$xg_{sat}(x,Q^{2})=\frac{16\mathcal{R}^{2}Q^{2}}{27{\pi}\alpha_{s}(Q^{2})}$.
The non-linear corrections to the longitudinal structure function
is defined as
\begin{eqnarray}
F^{NLC}_{L}(x,Q^{2})&=&C_{L,q}(\alpha_{s},x){\otimes}F_{2}^{s}(x,Q^{2})\\
&&+<e^{2}>C_{L,g}(\alpha_{s},x){\otimes}G^{NLC}(x,Q^{2}),\nonumber
\end{eqnarray}
Therefore the non-linear corrections to the reduced cross section
is defined by
\begin{eqnarray}
\sigma^{NLC}_{r}(x,Q^{2})&=&F_{2}(x,Q^{2})-f(y)F^{NLC}_{L}(x,Q^{2})
\end{eqnarray}
The analysis is performed in the ranges of
$10^{-5}{\leq}x{\leq}10^{-2}$ and
$1{\leq}Q^{2}{\leq}1000~\mathrm{GeV}^{2}$. The computed results of
the non-linear distribution  functions are compared with the
parametrization methods  [7,43] and the experimental data [46-48].\\
In Fig.1, the computed results of the non-linear gluon
distribution function are compared with the linear parametrization
model [7]. This behavior is considered at  $x<10^{-2}$ for
$Q^{2}=10, 30, 50$ and $100~\mathrm{GeV^{2}}$ in the hot-spot
point where the value of this parameter is defined to be
$R=2~\mathrm{GeV^{-1}}$ in this paper. In Fig.2 we show  the
non-linear results at an input $Q^{2}=1.9~\mathrm{GeV^{2}}$ in
comparison with the absorptive corrections at low $x$ and the
power corrections at low $Q^{2}$ values in Ref.[17]. As can be
observed in Fig.2, the behavior of the gluon distribution is flat
due to the freezing of $\alpha_{s}$ in comparison with the
non-linear behavior due to the parametrization model. The
confinement effect is expected to modify the running of the QCD
coupling $\alpha_{s}(Q^{2})$ to  $\alpha_{s}(Q^{2}+\mu_{0}^{2})$
where $\mu_{0}$ is the factorization scale. The results compared
with $\mu_{0}^{2}=0$, corresponding to no effects of confinement,
and with those obtained for $\mu_{0}^{2}=1~\mathrm{GeV}^{2}$ in Ref.[17].\\
In Fig.3 we make a critical study of the ratio $G/F_{2}$ proposed
in the last years [49,50] at linear and non-linear corrections,
which is frequently used to extract the gluon distribution from
the proton structure function. The ratio  $G/F_{2}$ is obtained at
$Q^{2}=5, 100$ and $1000~\mathrm{GeV^{2}}$ at low values of  $x$
with respect to the non-linear behavior of the gluon distribution
function due to the parametrization model. As can be observed in
Fig. 3, this ratio in a wide range of $x$ and $Q^{2}$ values is
dependent not only to $Q^{2}$, but also to $x$ at linear and
non-linear corrections. A purely $Q^{2}$ or $x$ independence of
the ratio were found in Refs.[50] to be not global in general as
compared with our results with respect to the parametrization
model.\\
H1 Collaboration [46] shows that measurement of the derivative
$(\partial{F_{2}}/\partial{\ln}Q^{2})_{x}$ has long been
recognised as a powerful constraint of the gluon density and
running coupling. For each bin of $x$, H1 Collaboration [46] shows
that these derivatives described by the function
$b(x)+2c(x){\ln}Q^{2}$, while in parametrization model it
described in terms of ${\ln}Q^{2}$ and ${\ln}x$ [7]. In Fig.4 the
linear and non-linear behavior of the quantity
$(\partial{F_{2}}/\partial{\ln}Q^{2})_{x}$ are considered and
compared with the H1 Collaboration data [46] as accompanied with
total errors. The non-linear correction to the derivative
$(\partial{F_{2}}/\partial{\ln}Q^{2})_{x}$ is performed due to the
non-linear gluon interaction effects in Eq.(2). The non-linear
behavior of the quantity is comparable with the H1 Collaboration
data in comparison with the linear behavior at low $x$. The
non-linear effects can be tested at a superior statistical accuracy attainable at the LHeC and FCC-he.\\
In the following, we  present the non-linear results that have
been obtained for the longitudinal structure function $F_{L}(x,
Q^{2})$, the ratio $F_{L}(x, Q^{2})/F_{2}(x, Q^{2})$ and the
reduced cross section $\sigma_{r}(x, Q^{2})$ from data mediated by
the parametrization of $F_{2}(x, Q^{2})$. The results for the
longitudinal structure function are presented in Fig.5 and
compared with the H1 data [48] as accompanied with total errors
where the average $x$ for each $Q^{2}$ is provided on the upper
scale of the figure. We use the non-linear longitudinal structure
function at NLO and NNLO approximations where effects of the
non-linear corrections to the gluon distribution are taken into
account. Non-linear results at NLO and NNLO approximations are
compared to the parametrization of $F_{L}$ at NLO approximation
[43] and also CT18 [4] at  NNLO approximation (CT18 results have
been performed at fixed value of the invariant mass $W$ as $W=230~
\mathrm{GeV}$). In Fig.6 the non-linear correction to the ratio
$F_{L}/F_{2}$ at  NNLO approximation is calculated and presented.
In this figure the ratio of the structure functions are compared
with the H1 Collaboration data [48]. The error bars of the ratio
$F_{L}/F_{2}$ are determined by
$\Delta({\frac{F_{L}}{F_{2}}})=\frac{F_{L}}{F_{2}}\sqrt{
({\frac{{\Delta}F_{L}}{F_{L}}})^{2}+({\frac{{\Delta}F_{2}}{F_{2}}})^2}$,
where ${\Delta}F_{L}$ and ${\Delta}F_{2}$ are collected from the
H1 experimental data in Ref.[48]. The non-linear results obtained
of the ratio $F_{L}/F_{2}$ are comparable to the results of the
color dipole model bounds [51] and experimental data [48]. This
comparisons are very good at low- and high-$Q^{2}$ values, even
compared to the parametrization model [7,43]. The good agreement
between the non-linear correction at NNLO analysis and the
experimental data indicates that these results have a bound
asymptotic behavior and they are compatible with the color dipole
model bounds. As can be observed in Fig. 6, the ratio has little
dependence on the $x$-evolution.\\
In Fig.7 we present the non-linear corrections to the reduced
cross section at NNLO approximation at $Q^{2}=8.5$ and
$18~\mathrm{GeV^{2}}$. As can be seen in this figure, one can
conclude that the non-linear corrections to the results
essentially improve the good agreement with  data  in comparison
with the parametrization model at low $Q^{2}$. HERA combined data
[47] are taken with center of mass energy
$\sqrt{s}=318~\mathrm{GeV}$ as accompanied with total errors.
These low-$x$ predictions are fully compatible with the H1 data
presented in [47]. The non-linear corrections are depicted and
compared with the linear parametrization models in this figure.
Consequently, the non-linear corrections make it possible to
perform the high-order corrections to the ultra-high-energy
processes.\\

\subsection{5. Summary}
In conclusion, we have studied the effects of adding the
non-linear corrections to the distribution functions for
transition from the linear to non-linear regions. We use the
parametrization of $F_{2}(x,Q^{2})$  and $F_{L}(x,Q^{2})$ as
baselines.  The non-linear corrections to the distribution
functions, to the derivative of the proton structure function, to
the ratio of structure functions, and to the reduced cross
sections at NNLO approximation are considered. Comparing these
quantities with the parametrization  and the color dipole models
indicates that the non-linear corrections  are enriched by the
behavior of distribution functions at low $Q^{2}$. The transition
of the ratio $F_{L}/F_{2}$ from the linear to the non-linear
behavior is considered and shows that it is in good agreement with
the color dipole model bounds not only at high-$Q^{2}$, but also
at low-$Q^{2}$ values. Comparison of the reduced cross sections
with respect to the non-linear  corrections with HERA data at low
and moderate $Q^{2}$ values shows that this transition has been
performed with good accuracy in comparison with the HERA combined
data. It has been found that at low and moderated $Q^{2}$, NNLO
results are corresponding to the experimental data and
parametrization methods. The non-linear method can be used in
 low-$x$ and low-$Q^{2}$ at the LHeC project. \\

%%%%%%%%%%%%%%%%%%%%%%%%%%%%%%%%%%%%%%%%%%%%%%%%%%%%%
\subsection{Appendix A}
The coefficient functions read as
\begin{eqnarray}
C_{2}&=&\widehat{A}_{2}+\frac{8}{3}a_{s}(Q^{2})DA_{2}\nonumber\\
C_{1}&=&\widehat{A}_{1}+\frac{1}{2}\widehat{A}_{2}+\frac{8}{3}a_{s}(Q^{2})D[A_{1}+(4\zeta_{2}-\frac{7}{2})A_{2}]\nonumber\\
C_{0}&=&\widehat{A}_{0}+\frac{1}{4}\widehat{A}_{2}-\frac{7}{8}\widehat{A}_{2}+\frac{8}{3}a_{s}(Q^{2})D[A_{0}
+(2\zeta_{2}-\frac{7}{4})A_{1}\nonumber\\
&&+(\zeta_{2}-4\zeta_{3}-\frac{17}{8})A_{2}],
\end{eqnarray}
\begin{eqnarray}
\widehat{A}_{2}&=&\widetilde{A}_{2}\nonumber\\
\widehat{A}_{1}&=&\widetilde{A}_{1}+2DA_{2}\frac{\mu^{2}}{\mu^{2}+Q^{2}}\nonumber\\
\widehat{A}_{0}&=&\widetilde{A}_{0}+DA_{1}\frac{\mu^{2}}{\mu^{2}+Q^{2}}\nonumber\\
\widetilde{A}_{i}&=&\widetilde{D}A_{i}+D\overline{A}_{i}\frac{Q^{2}}{Q^{2}+\mu^{2}}\nonumber\\
\widetilde{D}&=&\frac{M^{2}Q^{2}[(2-\lambda)Q^{2}+\lambda
M^{2}]}{[Q^{2}+M^{2}]^{3}}\nonumber\\
\overline{A}_{m}&=&a_{m1}+2a_{m2}L_{2},~~a_{02}=0.
\end{eqnarray}
and
\begin{eqnarray}
\widehat{B}^{(1)}_{L,s}&=&8C_{F}[\frac{25}{9}n_{f}-\frac{449}{72}C_{F}+(2C_{F}-C_{A})\nonumber\\
&&(\zeta_{3}+2\zeta_{2}-\frac{59}{72})]\nonumber\\
\overline{B}^{(1)}_{L,s}&=&\frac{20}{3}C_{F}(3C_{A}-2n_{f})\nonumber\\
\widehat{\delta}^{(1)}_{sg}&=&\frac{26}{3}C_{A}\nonumber\\
\overline{\delta}^{(1)}_{sg}&=&3C_{F}-\frac{347}{18}C_{A}\nonumber\\
\widehat{R}^{(1)}_{L,g}&=&-\frac{4}{3}C_{A}\nonumber\\
\overline{R}^{(1)}_{L,g}&=&-5C_{F}-\frac{4}{9}C_{A}\nonumber\\
L_{A}&=&L+\frac{A_{1}}{2A_{2}}\nonumber\\
L_{C}&=&L+\frac{C_{1}}{2C_{2}}\nonumber\\
L&=&\ln(1/x)+L_{1}\nonumber\\
L_{1}&=&{\ln}\frac{Q^{2}}{Q^{2}+\mu^{2}}\nonumber\\
L_{2}&=&{\ln}\frac{Q^{2}+\mu^{2}}{\mu^{2}}\nonumber\\
A_{i}(Q^{2})&=&\sum_{k=0}^{2}a_{ik}L_{2}^{k},~ (i=1,2)\nonumber\\
A_{0}&=&a_{00}+a_{01}L_{2}\nonumber\\
D&=&\frac{Q^{2}(Q^{2}+\lambda M^{2})}{(Q^{2}+M^{2})^2}.
\end{eqnarray}
%%%%%%%%%%%%%%%%%%%%%%%%%%%%%%%%%%%%%%%%%%%%%%%%%%%%%
\begin{table}[h]
\caption{ The effective parameters at low $x$ for
$0.11~\mathrm{GeV}^{2}<Q^{2}<1200~\mathrm{GeV}^{2}$ are defined by
the following values.}
\begin{tabular} {cccc}
\toprule \\  \multicolumn{2}{c}{parameters \quad \quad \quad ~~~~~~~~~~~~~~~~value}    \\ &&&\\ \hline \\ &&&\\
$a_{0}$& \quad  $-7.828\times 10^{-2}~\pm 5.19\times10^{-3}$ \\

$a_{1}$& \quad  $2.248\times 10^{-2}~\pm 1.47\times10^{-3}$\\

$a_{2} $  &   \quad  $2.301\times 10^{-4}~~  \pm
4.88\times10^{-4} $  \\&&&\\

$b_{0} $  &   \quad   $1.313\times 10^{-2}\pm 6.99\times10^{-4}$  \\

$b_{1}$   &    \quad  $4.736\times 10^{-3}\pm 2.98\times10^{-4}$ \\

$b_{2}$   &   \quad   $1.064\times 10^{-3}\pm 3.88\times10^{-5} $ \\&&& \\

$x_{P}$& \quad  $0.0494~\pm 0.0039$ \\

$F_{P}$& \quad  $0.503~\pm 0.012$ \\

$\chi^{2}_{min}$ &  \quad  $193.19$ \\

\hline

\end{tabular}
\end{table}
\begin{table}[h]
\caption{ The effective parameters used in appendix A with respect
to Eqs.(9) and (10) in Ref.[43].}
\begin{tabular} {cccc}
\toprule \\  \multicolumn{2}{c}{parameters \quad \quad \quad ~~~~~~~~~~~~~~~~value}    \\ &&&\\ \hline \\ &&&\\
$a_{00}$& \quad  $2.550\times 10^{-1}~\pm 1.600\times10^{-2}$ & &\\
$a_{01}$& \quad  $1.475\times 10^{-1}~\pm 3.025\times10^{-2}$ \\& &\\

  $a_{10} $  &   \quad  $8.205\times 10^{-4}~~  \pm  4.62\times10^{-4} $  \\

  $a_{11} $  &   \quad   $-5.148\times 10^{-2}\pm 8.19\times10^{-3}$  \\

  $a_{12}$   &    \quad  $-4.725\times 10^{-3}\pm 1.01\times10^{-3}$   \\  &&&\\

 $a_{20}$   &   \quad   $2.217\times 10^{-3}\pm 1.42\times10^{-4} $ \\

 $a_{21}$   &   \quad   $1.244\times 10^{-2}\pm 8.56\times10^{-4}$  \\

 $a_{22}$    &    \quad  $5.958\times 10^{-4}\pm 2.32\times10^{-4} $ \\ &&& \\

\hline

\end{tabular}
\end{table}
%%%%%%%%%%%%%%%%%%%%%%

%%%%%%%%%%%%%%%%%%%%%%%%%%%%%%%%%%%%%%%%%%%%%%%%%%%%%%

\subsection{ACKNOWLEDGMENTS}

We are grateful to the Razi University for financial support of
this project. G.R.Boroun was especially grateful to A.V.Kotikov
for
carefully reading the paper and for critical notes.\\

%%%%%%%%%%%%%%%%%%%%%%%%%%%%%%%%%%%%%%%%%%%%%%%%%%%%%%%%%%%%%%%%%%%%%%%%
%%%%%%%%%%%%%%%%%%%%%%%%%%%%%%%%%%%%%%%%%%%%%%%%%%%%%

%%%%%%%%%%%%%%%%%%%%%%%%%%%%%%%%%%%%%%%%%%%%%%%%%%%
\section{References}
1. L.N. Lipatov, Sov. J. Nucl. Phys.{\bf20}, 94 (1975); V.N.
Gribov, L.N. Lipatov, Sov. J. Nucl. Phys.{\bf15}, 438 (1972); G.
Altarelli, G. Parisi, Nucl. Phys. B{\bf126}, 298 (1977); Yu.L.
Dokshitzer, Sov. Phys.
JETP {\bf46}, 641 (1977).\\
2. A.D.Martin, R.G.Roberts, W.J.Stirling and R.S.Thorne,
Eur.Phys.J.C{\bf23}, 73(2002); Phys.Lett.B {\bf531}, 216 (2002).\\
3. CTEQ Collaboration, J.Pumplin et al., J.High Energ.Phys.{\bf07}, 012 (2002).\\
4. Tie-Jiun Hou et al., Phys.Rev.D {\bf103}, 014013 (2021).\\
5. H1 Collaboration, C.Adloff et al., Eur.Phys.J.C {\bf12},
375 (2000).\\
6. M. Froissart, Phys.Rev.{\bf123}, 1053 (1961).\\
7.  M. M. Block and L. Durand,  arXiv [hep-ph]: 0902.0372
(2009).\\
8. M. M. Block, L. Durand and P. Ha, Phys. Rev.D {\bf 89}, 094027 (2014).\\
9. R. Fiorea et al., Phys.Rev.D {\bf71}, 033002 (2005); R.Fiorea et al., Phys.Rev.D {\bf73}, 053012 (2006).\\
10. LHeC Collaboration and FCC-he Study Group
, P.Agostini et al., CERN-ACC-Note-2020-0002, arXiv:2007.14491 [hep-ex] (2020).\\
11. M.Klein, Annalen Phys.{\bf528}, 138 (2016); M.Klein,  arXiv
[hep-ph]:1802.04317.\\
12. N.Armesto et al.,
Phys.Rev.D {\bf100}, 074022 (2019).\\
13. K.J.Eskola et al., Nucl.Phys.B {\bf660}, 211 (2003);
K.J.Eskola et al., arXiv: hep-ph/0302185 (2003); M.A.Kimber,
J.Kwiecinski and A.D.Martin, Phys.Lett.B {\bf508}, 58 (2001).\\
14. K.Prytz, Eur.Phys.J.C {\bf22}, 317 (2001).\\
15. G.Ingelman and K.Prytz, Z.Phys.C {\bf58}, 285 (1993).\\
16. B.Rezaei and G.R.Boroun, Phys.Lett.B {\bf692}, 247 (2010).\\
17. M.R.Pelicer et al., Eur.Phys.J.C {\bf79}, 9 (2019).\\
18. G.R.Boroun, Eur.Phys.J.A {\bf43}, 335 (2010).\\
19. M.Devee and J.K.Sarma, Eur.Phys.J.C {\bf74}, 2751 (2014);
 arXiv [hep-ph]: 1808.01586 (2018); Nucl.Phys.B {\bf885}, 571 (2014); M.Devee, arXiv [hep-ph]: 1808.00899 (2018).\\
20. B.Rezaei and G.R.Boroun, Phys.Rev.C {\bf101}, 045202 (2020).\\
21. M.Lalung, P.Phukan and J.K.Sarma, Int.J.Theor.Phys.{\bf56},
11(2017); Nucl.Phys.A {\bf992}, 12615 (2019); arXiv [hep-ph]:1801.06360(2019).\\
22. G.R.Boroun, Phys.Rev.C {\bf97}, 015206 (2018).\\
23. H.Khanpour, Phys.Rev.D {\bf99}, 054007 (2019).\\
24. G.R.Boroun and S.Zarrin, Eur.Phys.J.Plus {\bf128}, 119 (2013).\\
25. P.Phukan, M.Lalung and J.K.Sarma, Nucl.Phys.A {\bf968},
275 (2017).\\
26. B.Rezaei and G.R.Boroun, Eur.Phys.J.A {\bf55}, 66 (2019).\\
27.R.Wang and X.Chen, Chinese Phys.C {\bf41}, 053103 (2017).\\
28. G.R.Boroun and B.Rezaei, Nucl.Phys.A {\bf1006}, 122062
(2021).\\
29. A.Kovner and Urs A.Wiedemann, Phys.Rev.D {\bf66},
051502 (2002).\\
30. G.R.Boroun, JETP Letters {\bf114}, 1 (2021).\\
31. J.Gao, L.Harland-Lang and J.Rojo,
Phys.Rept.{\bf742}, 1 (2018).\\
32. I.I.Balitsky and A.V.Belitsky, Nucl.Phys.B {\bf629}, 290
(2002).\\
33. A.M.Stasto, Acta Phys.Polon.B {\bf33}, 1571(2002).\\
34. R.D.Ball et al., Eur.Phys.J.C {\bf78}, 321 (2018);
 xFitter Collaboration, H.Abdolmaleki et al.,
 arXiv:1802.00064.\\
35. J.Bartels and E.Levin, Nucl.Phys.B {\bf387}, 617 (1992).\\
36. L.V.Gribov, E.M.Levin and M.G.Ryskin, Phys.Rept.{\bf100}, 1
(1983); A.H.Mueller and J.w.Qiu, Nucl.Phys.B {\bf268},
427 (1986).\\
37. J.Kwiecinski et al., Phys.Rev.D {\bf42}, 3645 (1990).\\
38. J. Blumlein, V. Ravindran and W. van Neerven, Nucl. Phys. B
\textbf{586}, 349 (2000); S.Catani and F.Hautmann,
Nucl.Phys.B {\bf427}, 475 (1994).\\
39. D.I.Kazakov and A.V.Kotikov, Phys.Lett.B {\bf291}, 171 (1992);
E.B.Zijlstra and W.L.van Neerven, Nucl.Phys.B {383}, 525 (1992).\\
40. W.L. van Neerven, A.Vogt, Phys.Lett.B \textbf{490}, 111 (2000); A.Vogt, S.Moch, J.A.M.Vermaseren, Nucl.Phys.B \textbf{691}, 129 (2004).\\
41. G.Altarelli and G.Martinelli, Phys.Lett.B \textbf{76}, 89 (1978).\\
42. S.Moch, J.A.M.Vermaseren, A.Vogt, Phys.Lett.B \textbf{606},
123 (2005).\\
43. L.P.Kaptari et al., Phys.Rev.D{\bf 99}, 096019 (2019); L.P.Kaptari et al., JETP Lett.{\bf 109}, 281 (2019)\\
44. B.Rezaei and G.R.Boroun, Eur.Phys.J.A {\bf56}, 262 (2020).\\
45. A.V.Kotikov, JETP Lett.{\bf111}, 67 (2020).\\
46. H1 Collaboration, C.Adloff et al., Eur.Phys.J.C {\bf21},
33 (2001).\\
47. H1 Collaboration and ZEUS Collaboration, H. Abramowicz et al.,
 Eur. Phys. J. C {\bf75}, 580 (2015).\\
48. H1 Collaboration, V. Andreev et al., Eur. Phys. J. C {\bf74},
2814 (2014).\\
49. G.R.Boroun, Eur.Phys.J.A {\bf50}, 69 (2014).\\
50. D.K.Choudhury and L.Machahari, arXiv[hep-ph]:2007.00978;
L.Machahari and D.K.Choudhury, Eur.Phys.J.A {\bf54}, 69 (2018);
J.K.Sarma, K.Choudhury and G.K.Medhi, Phys.Lett.B {\bf403}, 139
(1997);
 M.Devee, R. Baishya and J.K.Sarma, Eur.Phys.J.C {\bf72}, 2036 (2012).\\
51. C. Ewerz et al., Phys.lett.B {\bf720}, 181 (2013).\\

%%%%%%%%%%%%%%%%%%%%%%%%%%%%%%%%%%%%%%%%%%%%%%%%
\newpage
\begin{figure}
\includegraphics[width=1\textwidth]{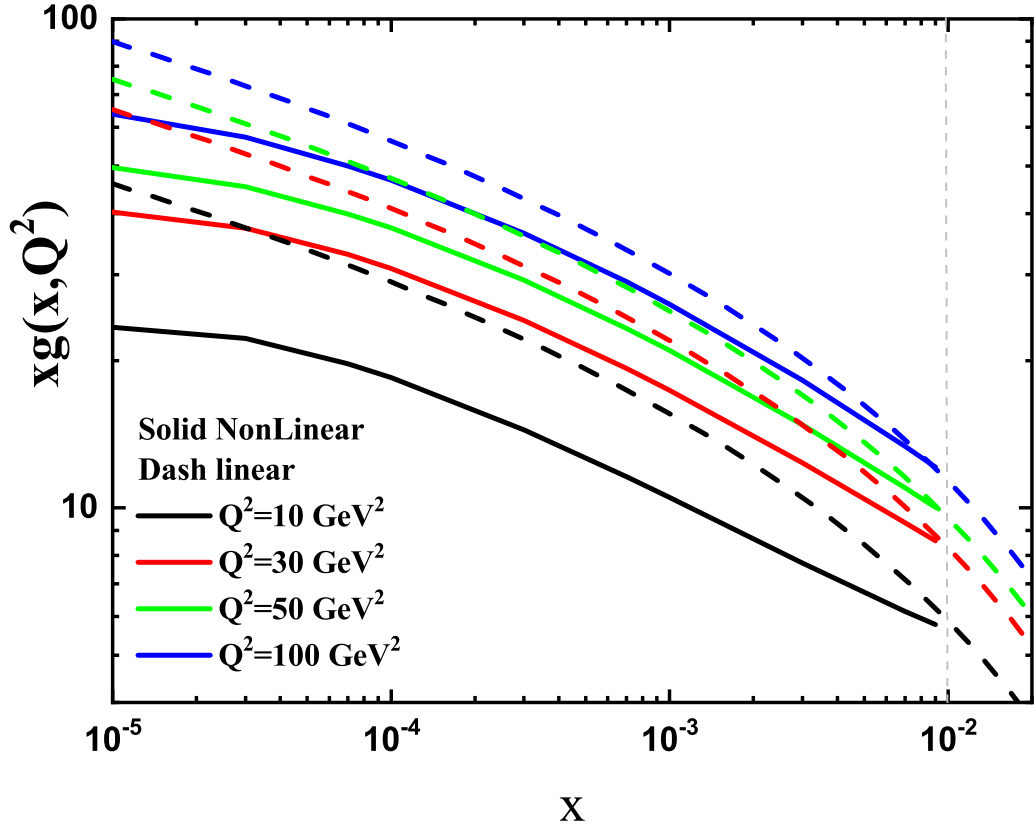}
\caption{The linear and non-linear gluon distribution function at
$R=2~\mathrm{GeV}^{-1}$ for $Q^{2}=10, 30, 50$ and
$100~\mathrm{GeV^{2}}$ with respect to the parametrization model
[7] and GLR-MQ equation [36] respectively.}\label{Fig1}
\end{figure}
\begin{figure}
\includegraphics[width=1\textwidth]{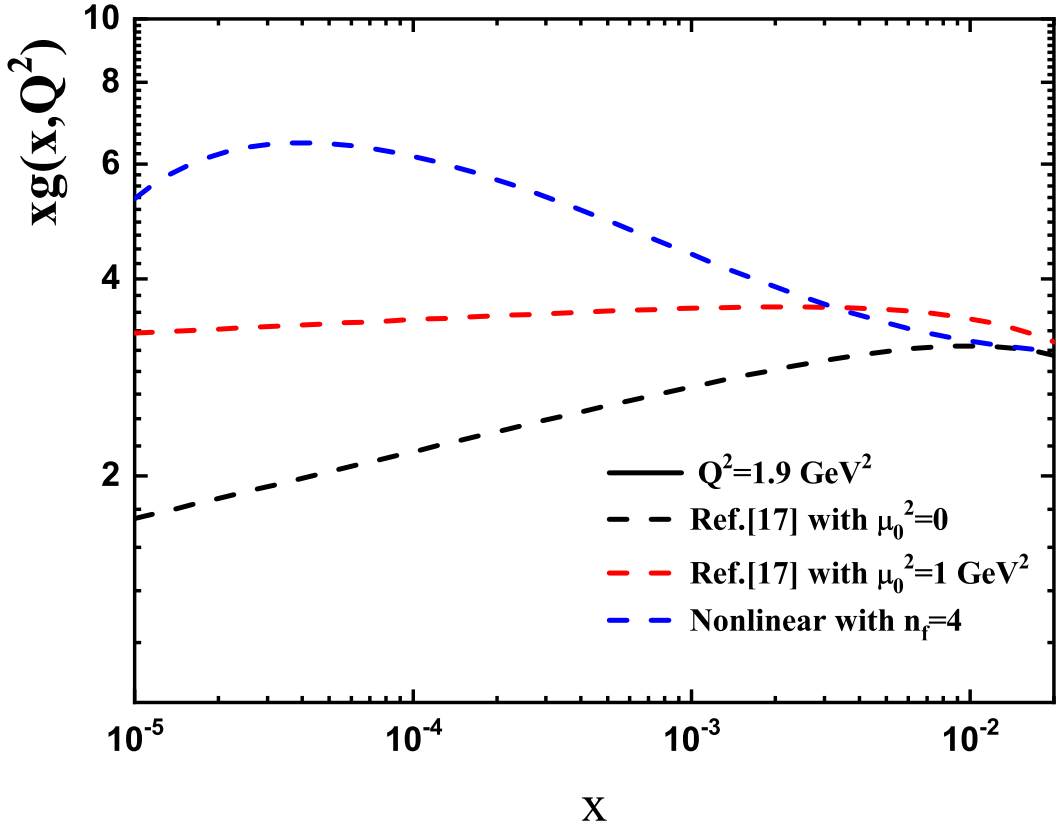}
\caption{The non-linear gluon distribution function at
$R=2~\mathrm{GeV}^{-1}$ and $n_{f}=4$ at
$Q^{2}=1.9~\mathrm{GeV^{2}}$ compared with the power corrections
arises from the freezing of the running coupling  with
$\mu_{0}^{2}=0$ and $1~\mathrm{GeV^{2}}$ [17].}\label{Fig2}
\end{figure}
\begin{figure}
\includegraphics[width=1\textwidth]{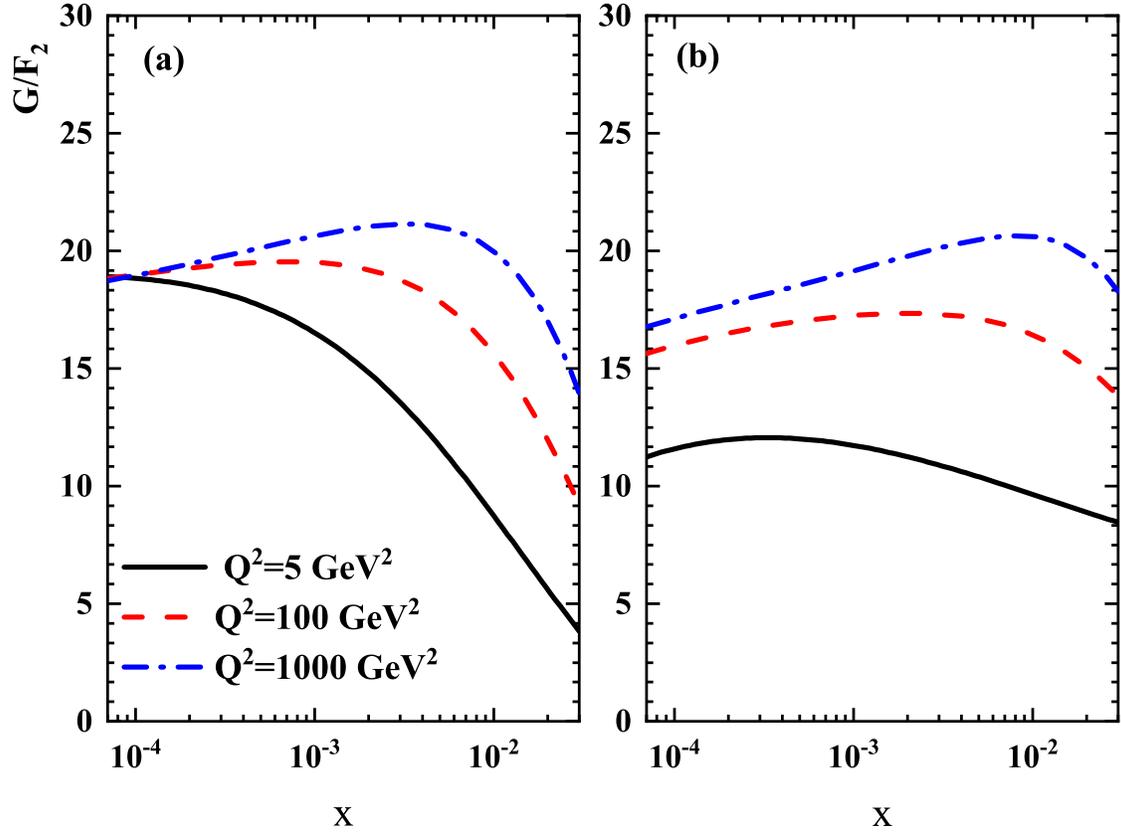}
\caption{Results of the ratio $G(x,Q^{2})/F_{2}(x,Q^{2})$ at
$Q^{2}=5, 100$ and $1000~\mathrm{GeV^{2}}$ vs $x$ obtained from
(a) the linear  and (b) the non-linear  parametrization
model.}\label{Fig3}
\end{figure}
\begin{figure}
\includegraphics[width=1\textwidth]{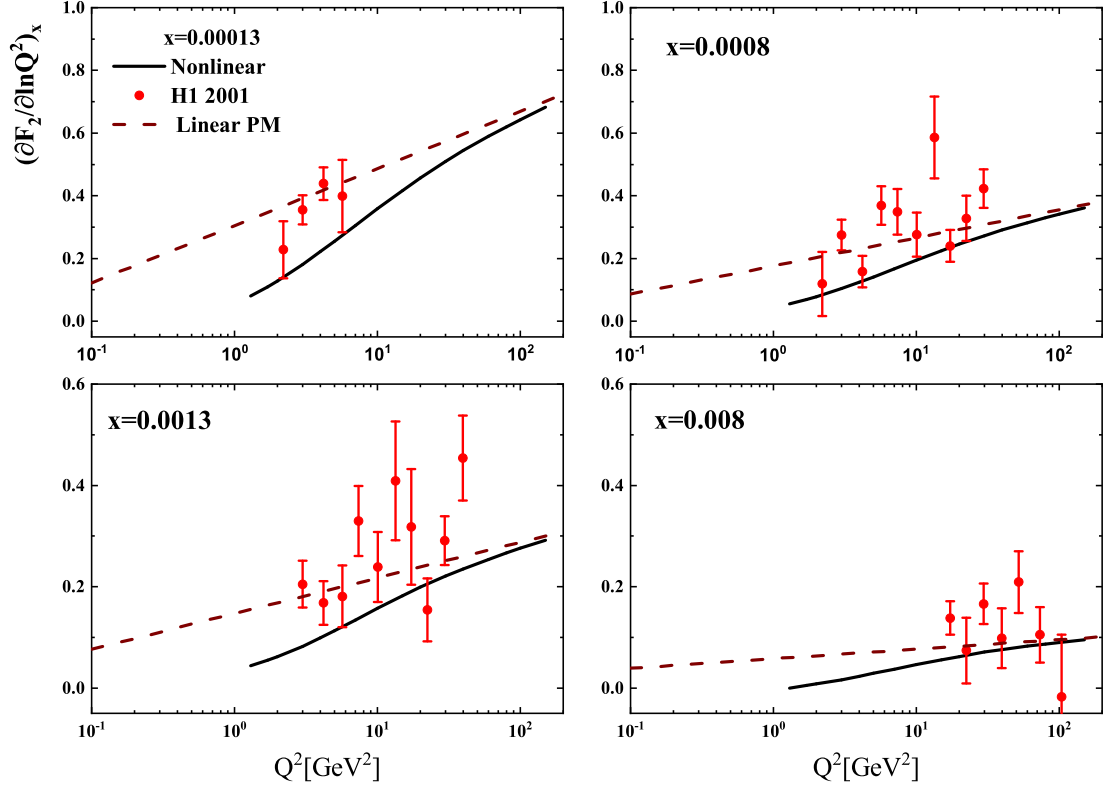}
\caption{The derivative $(\partial{F_{2}}/\partial{\ln}Q^{2})_{x}$
plotted as functions of $Q^{2}$ for fixed $x$ compared with  the
H1 Collaboration data [46] as accompanied with total errors. The
linear and non-linear behaviors are obtained from the
parametrization model.}\label{Fig4}
\end{figure}
\begin{figure}
\includegraphics[width=1\textwidth]{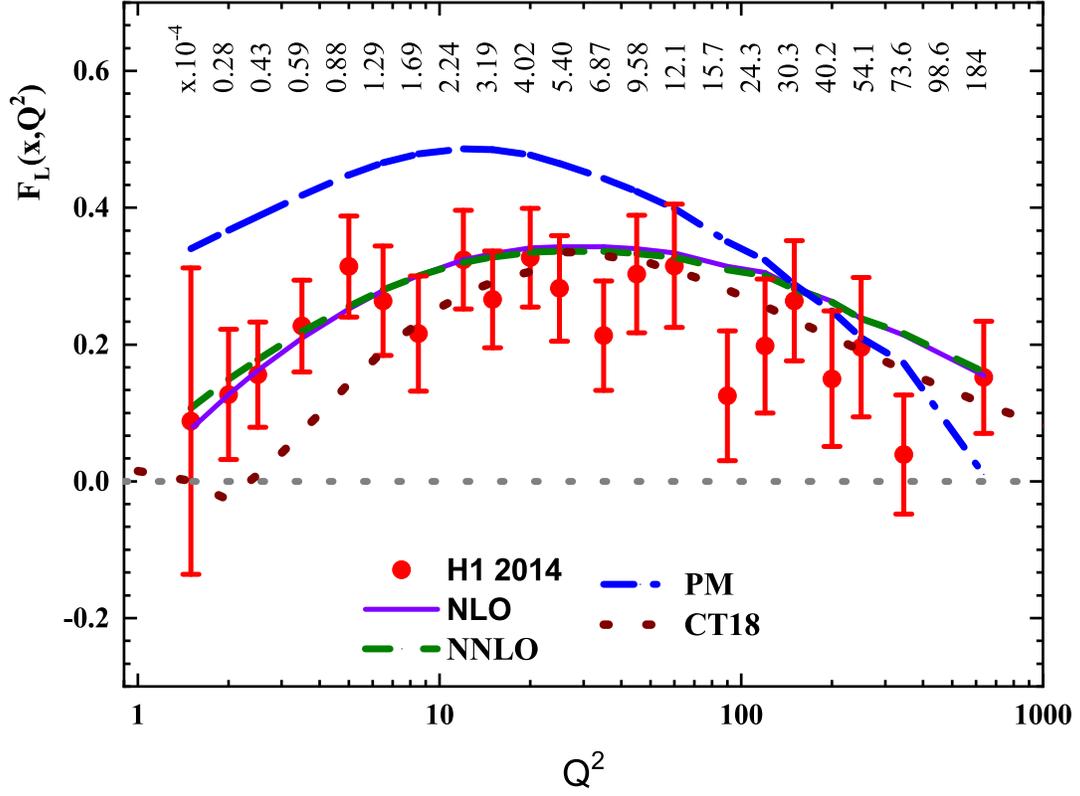}
\caption{The non-linear longitudinal structure function $F_{L}$ at
NLO and NNLO approximations averaged over $x$ at different
$Q^{2}$. The average value of $x$ for each $Q^{2}$ is given above
each data point. The non-linear results at NLO (solid) and NNLO
(dashed) are compared to the H1 Collaboration data [48] as
accompanied with total errors, the parametrization model [43]
(dashed-dot) and CT18 [4](dot) at the NNLO approximation at fixed
value of the invariant mass $W=230~ \mathrm{GeV}$.}\label{Fig5}
\end{figure}
\begin{figure}
\includegraphics[width=1\textwidth]{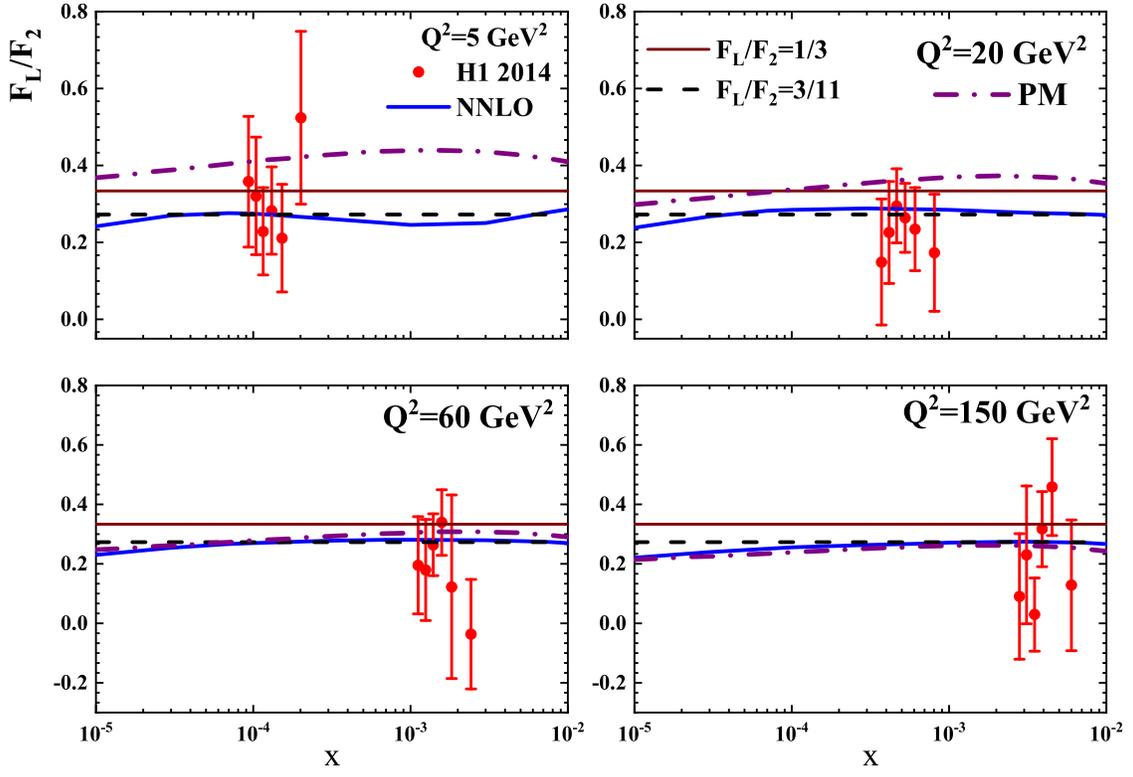}
\caption{Results of the ratio $F_{L}/F_{2}$ obtained from the
 non-linear  corrections at  NNLO approximation at fixed $Q^{2}$ values.
 The ratio compared with the color dipole picture bounds
  (i.e., $F_{L}/F_{2}=1/3$ and $3/11$), H1 Collaboration data [48] as accompained with
  total errors, and the parametrization model [7,43].}\label{Fig6}
\end{figure}
\begin{figure}
\includegraphics[width=1\textwidth]{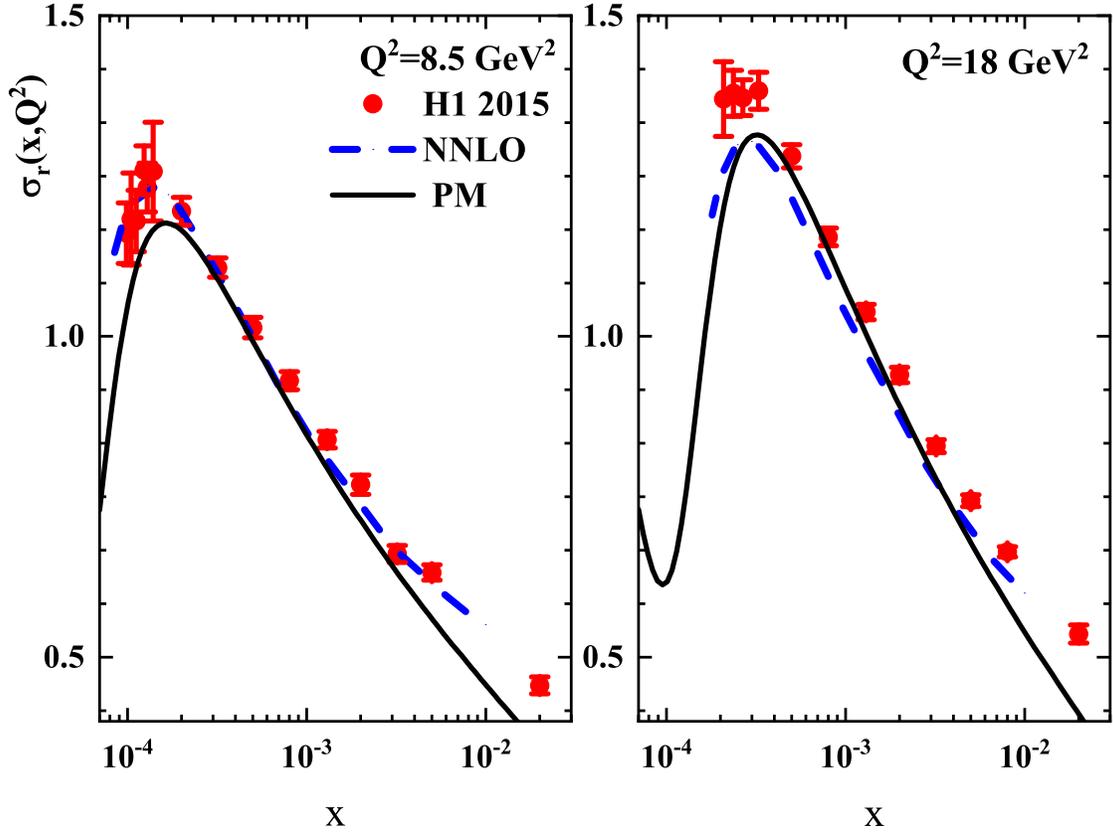}
\caption{The  NNLO predictions for the reduced DIS cross section
at $Q^{2}=8.5$ and $18~\mathrm{GeV^{2}}$ available. The H1 data
for some representative fixed values of $Q^{2}$ are taken from
[47] as accompanied with total errors. The non-linear corrections
of the reduced cross section at NNLO approximation compared with
the parametrization model.}\label{Fig7}
\end{figure}

%%%%%%%%%%%%%%%%%%%%%%%%%%%%%%%%%%%%%%%%%%%%%%%%%

%%%%%%%%%%%%%%%%%%%%%%%%%%%%%%%%%%%%%%%%%%%%%%%%%%%%%%%
%%%%%%%%%%%%%%%%%%%%%%%%%%%%%%%%%%%%%%%%%%%%%%%%%%%%%%%%
\end{document}